\documentclass{iaus}
\usepackage{graphicx}

\title[Simulations of Core Convection] 
{Simulations of core convection and resulting dynamo action in rotating
A-type stars}

\author[Browning, Brun \& Toomre]   
{Matthew K. Browning$^1$,
 Allan Sacha Brun$^{2,1}$ \break \and Juri Toomre$^1$}

\affiliation{$^1$JILA, University of Colorado at Boulder, 440 UCB, Boulder CO, USA
80309-0440 \break email: matthew.browning@colorado.edu\\[\affilskip]
$^2$SAp, CEA-Saclay, 91191 Gif-sur-Yvette, France}

\pubyear{2004}
\volume{224}  
\pagerange{119--126}
\date{?? and in revised form ??}
\jname{The A-Star Puzzle}
\editors{J.\,Zverko, W.W.\,Weiss, J.\,\v{Z}i\v{z}\v{n}ovsk\'{y}, \& S.J.\,Adelman, eds.}
\begin{document}

\maketitle

\begin{abstract}
We present the results of 3--D nonlinear simulations of magnetic dynamo
action by core convection within A-type stars of 2 solar masses, at a range
of rotation rates.  We consider the inner 30\% by radius of such stars, with
the spherical domain thereby encompassing the convective core and a portion
of the surrounding radiative envelope.  The compressible Navier-Stokes
equations, subject to the anelastic approximation, are solved to examine
highly nonlinear flows that span multiple scale heights, exhibit intricate
time dependence, and admit magnetic dynamo action.  Small initial seed
magnetic fields are found to be amplified greatly by the convective and
zonal flows.  The central columns of strikingly slow rotation found in some
of our progenitor hydrodynamic simulations continue to be realized in some
simulations to a lesser degree, with such differential rotation arising
from the redistribution of angular momentum by the nonlinear convection and
magnetic fields.  We assess the properties of the magnetic fields thus
generated, the extent of convective penetration, the magnitude of the
differential rotation, and the excitation of gravity waves within the
radiative envelope.
\keywords{convection, magnetic fields, turbulence, stars: interiors}
\end{abstract}

\firstsection 
\section{Introduction}

The observational pathologies of A-type stars have attracted scrutiny for
more than a century (e.g., Maury 1897). The striking surface features seen
in some of these stars are varied and extensive: abundance anomalies,
strong global magnetic fields, rapid acoustic oscillations.  Though many of
these may be driven primarily by phenomena near the stellar surface, some
could be influenced by the core convection occurring deep within the
interiors of A-type stars.  In particular, the long-standing puzzle
regarding the origins of the surface magnetism in Ap stars raises questions
about whether that core convection can sustain dynamo action, and if so,
whether such interior fields could have any influences at the stellar
surface or are instead buried from view.

The convection within the core clearly impacts the structure and evolution
of massive stars.  Convective motions can overshoot beyond the region of
superadiabatic stratification, bringing fresh fuel into the nuclear-burning
core and prolonging the star's main-sequence lifetime.  Observational hints
at the extent of these overshooting motions have been provided through
studies of best-fit isochrones to clusters (e.g., Meynet, Mermilliod \&
Maeder 1993).  Likewise some theoretical estimates have served to constrain
the penetrative properties of the convection (Roxburgh 1978, 1989, 1992).
Yet a detailed model of the overshooting realized from convective cores --
one that gives its extent and its variation with latitude -- remains
lacking.

Motivated in part by these observational and theoretical challenges, we
have conducted 3-D simulations of core convection deep within A-type stars.
We begin by discussing first some properties of such non-magnetic modelling from
Browning, Brun, \& Toomre (2004), and then turn to our most recent MHD
simulations.  Among our aims here are to assess the penetration and
overshooting achieved from convective cores, and the nature of the
convective flows and differential rotation that are established.  In the calculations with
magnetism, we explore whether dynamo action is realized, and if so what are
the gross properties of the magnetism -- its morphology, its intensity, its
variations with time.  Although the wealth of A-star observations lends
vibrancy to our theoretical study, our work has little to say about the
surface conundrums at this stage.  The models here simply provide some
glimpses into the complicated dynamics happening within the central regions
of these stars.

\section{Approach}
Our simulations here examine the inner 30\% by
radius of A-type stars of two solar masses, rotating at the solar rate
(case A) and fourfold faster (case B). The spherical domain thus
encompasses the entire convective core (central 15\% of star), but only a
portion of the surrounding radiative envelope.  We exclude the innermost
2\% of the star from our computations for numerical reasons.  Thus we
consider turbulent convection interacting with rotation and magnetism
mainly within the cores of such stars.  We employ our Anelastic Spherical Harmonic
(ASH) code, which solves the full 3-D anelastic MHD equations (Brun,
Miesch \& Toomre 2004).  The radial stratification is consistent with a
1-D stellar structure model.  The MHD simulations were begun by introducing
small seed dipole magnetic fields into progenitor mature non-magnetic
simulations (Browning, Brun \& Toomre 2004).  We adopt a Prandtl number
$P_r$ of 0.25, a magnetic Prandtl number $P_m$ of 5, and consider
stratifications within the radiative envelope that are somewhat less stable
(subadiabatic) than those of real stars.  We adopt stress-free and
impenetrable boundary conditions at the top and bottom of the domain, and
require the magnetic field to be purely radial there.  The intricate
spatial variation of flows and magnetic fields in our simulations is
captured by expanding both in spherical harmonics in the horizontal
directions (including all degrees up to $\ell=170$) and in Chebyshev
polynomials in the radial (employing two stacked expansion domains with
$N_r$=82 colocation points).  The high-resolution modelling of the evolving
turbulent convection and intricate magnetism requires the use of massively parallel
supercomputers.

\section{Results}

\begin{figure}
\includegraphics[width=4.5in]{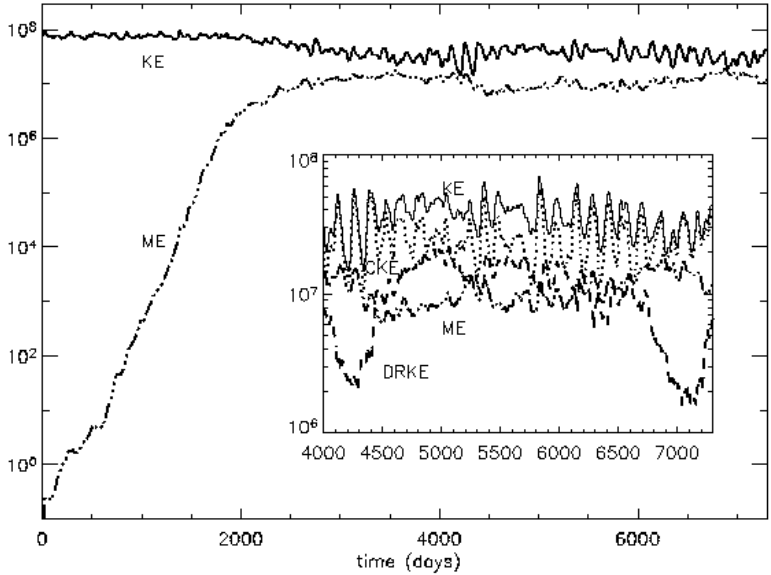}
\caption{Temporal evolution of volume-averaged magnetic energy density (ME)
from an initial seed dipole field, together with total kinetic energy
density (KE).  \textit{Inset}: vigorous time dependence of the energy densities in
the convection (CKE) and differential rotation (DRKE), with pronounced
successive minima in DRKE.}
\end{figure}

Our simulations without magnetism (Browning et al. 2004) revealed that the
convective flows within the core are vigorous and time dependent.  The
global connectivity allowed in spherical domains means that the
motions can couple widely separated sites, extend radially through much of
the core, and span large fractions of a hemisphere.  The flows
penetrated into the surrounding radiative envelope, establishing a
nearly-adiabatic penetrative region that was prolate in shape, having
greater spatial extent near the poles than at the equator.  The farther
region of overshooting, where flows persisted but did not appreciably modify
the prevailing stratification, had a basically spherical outer boundary.
By pummelling the lower boundary of the radiative envelope, the penetrative
convective plumes excited internal gravity waves within the radiative
envelope.

The present MHD simulations also reveal that sustained dynamo action is
realized, with the vigorous core convection serving to amplify initial seed
fields by many orders of magnitude (Fig. 1).  The magnetic fields that
result ultimately possess energy densities comparable to that in the flows
relative to the rotating frame.  The magnetism is then sustained against
decay for the multiple ohmic diffusion times that we have studied.  Both
simulations exhibit rich time dependence, as magnetic fields are generated
by the convective and zonal flows and in turn feed back upon the motions,
yielding the pronounced oscillatory behavior in the timetrace of energy
densities (Fig. 1).  

The morphologies of the flows and magnetic fields within the core are
distinctive (Fig. 2): the radial velocity ($v_r$) shows columnar alignment
reminiscent of Taylor columns, whereas the magnetism is both fibril ($B_r$)
and stretched into ribbon-like structures that extend around much of the
core ($B_{\phi}$).  The differing morphologies of $v_r$ and $B_r$ arise
partly from the smaller value of magnetic diffusivity relative to the
viscous diffusivity.  The organized bands of toroidal field may be
attributable to stretching by angular velocity gradients, described in
mean-field dynamo theory as the $\omega$-effect, near the boundary between
the core and the radiative envelope.  Such stretching and amplification of
the field by differential rotation also means that $B_{\phi}$ near the core
boundary is only slightly diminished from its interior strength.  In
contrast, $v_r$ and $B_r$ decline rapidly in amplitude outside the
convective core (Fig. 2$a$,$b$).  Within the core, $B_r$ and $B_{\phi}$
possess comparable strengths, but in the region of overshooting, where
convective motions have waned, $B_r$ has decreased by about a factor of 40
whereas $B_{\phi}$ is only roughly a factor of 3 smaller than in the core.

The magnetic fields realized within the core are predominantly fluctuating,
accompanied by weak mean fields.  Throughout the core, the energy
associated with fluctuating fields accounts for more than 90\% of the total
magnetic energy, split in roughly equal measure between toroidal and
poloidal fluctuating components.  In contrast, the comparatively weak mean
(axisymmetric) fields are divided unevenly between toroidal and poloidal
components, with $B_{t}$ stronger than $B_{p}$ by about a
factor of two.  Outside the core, the mean toroidal field becomes the
dominant component of the plummeting magnetic energy.  This likely arises
because the generation of fields by helical convection largely vanishes,
but the stretching and amplification of toroidal fields by angular velocity
contrasts persists.

The strong differential rotation established in simulations without
magnetic fields by the Reynolds stresses is diminished by the presence of
magnetism (Fig. 3) due to the poleward transport of angular momentum by
Maxwell stresses.  In case A, with ME about 40\% of KE, prominent
differential rotation continues to be realized. However, in the more
rapidly rotating case B with nearly equipartition fields, contrasts in
angular velocity are greatly weakened.  Case A exhibits major variations in
differential rotation, with brief intervals during which the angular
velocity contrasts become quite small.  These grand minima in the
differential rotation are visible in Figure 1 (\textit{inset}) as broad dips in DKRE.  The
onsets of these intervals of small DRKE coincide with times when ME has
grown larger than about 40\% of KE, suggesting a complex interplay between
strengthening magnetic fields and weakening differential rotation.

\begin{figure}
\includegraphics[width=4.9in]{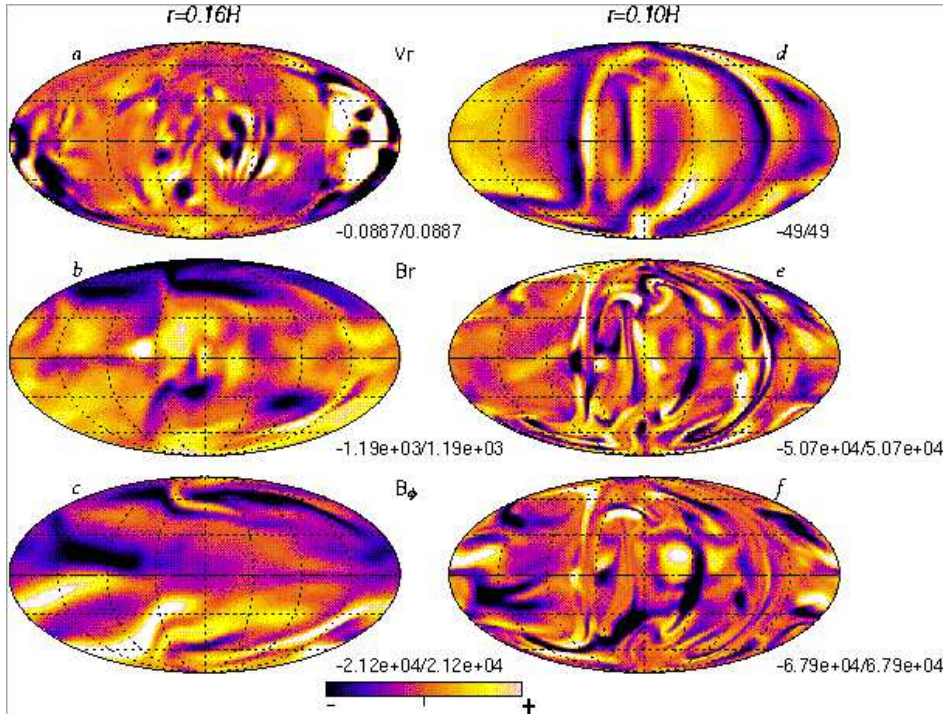}
\caption{Instantaneous global views for case A of the fields $v_r$, $B_r$, and
$B_{\phi}$ on spherical surfaces (rendered as Mollweide projections) at mid-core ($r=0.10 R$, right), and in
the region of overshooting ($r=0.16 R$, left).  Dashed line denotes
equator.  Positive values are shown in bright tones, negative values in
dark tones, and the ranges rendered are indicated in m\,s$^{-1}$ and in
G.  Within the core, intricate magnetic fields are well-correlated with the
broad convective flows, whereas outside, differential rotation stretches
$B_{\phi}$ into larger-scale features.} 
\end{figure}

\begin{figure}
\includegraphics[width=4.5in,trim=0 0 0 0]{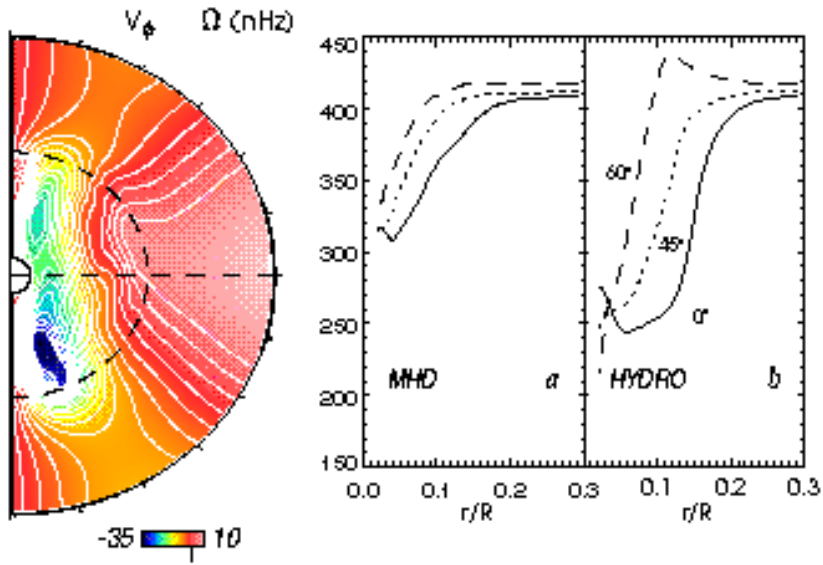}
\caption{Differential rotation realized in case A and its hydrodynamic
progenitor.  (\textit{left}) Contour plot in radius and latitude of
longitudinal (zonal) velocity $\hat{v}_{\phi}$, with color bar and ranges
(in m\,s$^{-1}$) indicated.  (\textit{right}) Angular velocity
$\hat{\Omega}$ with radius for indicated latitudinal cuts in both ($a$)
case A and ($b$) its progenitor.  The magnetism from dynamo action weakens
the strong angular velocity contrasts seen in the hydrodynamic
simulations.}
\end{figure}

\section{Reflections}

Our simulations have revealed some striking dynamical properties of core
convection deep within prototypical A-type stars.  The vigorous flows
penetrate into the surrounding radiative envelope, thereby exciting gravity
waves, and can establish differential rotation within the core.  Our MHD
models further show that the flows admit magnetic dynamo action, with
initial seed fields amplified by many orders of magnitude and sustained
against ohmic decay.  The resulting magnetic fields are mostly
non-axisymmetric and intermittent in nature, and attain strengths of up to
several hundred kG.

The generation of magnetic fields in our simulations is a complex process.
Both helical convective motions and differential rotation appear to play
roles in giving rise to the mean fields within the core.  Thus the dynamo
operating there may loosely be classified in the language of mean-field
theory as being of the $\alpha^2-\omega$ type. The generation of the far
stronger fluctuating fields is less readily parameterized in terms of such
concepts, but there, too, convection and differential rotation are likely
both important.  Whether such differential rotation is indeed an essential
element in the operation of the dynamo, and how the balance between it and
the magnetism is set, remains unclear at this stage.

How the magnetic fields and flows realized in our simulations are affected
by rotation, and how they are influenced by our simulation parameters, are
both likely to be sensitive matters.  Future work is required to assess
whether the dynamos operating here are realized at all rotation rates, and
if so, how the strength of the fields may scale with rotation.  Also
unaddressed by the present simulations is the question of whether the
dynamo-generated fields can migrate to the surface where they might be
observed.  Magnetic buoyancy instabilities could potentially lead to such
field emergence (e.g., MacGregor \& Cassinelli 2003) given strong enough
magnetism, but MacDonald \& Mullan (2004) indicate difficulties with this
process for realistic stratifications.  The highly fibril interior fields
needed for such instabilities to arise are also somewhat at odds with the
large-scale magnetism observed at the surface, though the detailed
morphology of fields that have risen from the core through the radiative
envelope is difficult to predict.

\begin{acknowledgments}
We thank Douglas Gough for helpful discussions, NASA for providing partial
funding through SEC Theory Program grant NAG5-11879 and through the
Graduate Student Researchers Program (NGT5-50416), and NSF PACI support for
providing supercomputing resources at PSC, NCSA, and SDSC. Browning's
travel was supported by the AAS and NSF through an International Travel
Grant.
\end{acknowledgments}

\begin{discussion}
\discuss{Dworetsky}{For many years the generally accepted model for Ap star
  fields has been the ``frozen fossil field.''  Does this work imply that
  the dynamo model for Ap stars is staging a comeback?  Do I need to change
  my standard ``Ap star'' lecture slides?}

\discuss{Browning}{No, I'm not advocating a change in worldview regarding the
  magnetism of Ap stars.  We've shown that strong fields are produced
  within the core, but getting those fields to the surface is another
  matter entirely.  At this stage, we simply don't know for sure whether
  that's possible; modelling by MacGregor and Cassinelli, and more recently
  MacDonald \& Mullan, though, suggests that it's very difficult to get the
  fields out.}

\discuss{Piskunov}{Do you see a magnetic field strength maximimum at some
  spherical harmonics and if yes, what order of spherical harmonic is it?}

\discuss{Browning}{The power spectra are fairly flat until about $\ell=15$,
  after which they fall off rapidly.  There's a modest peak at around
  $\ell=7$.}

\discuss{Moss}{First, a comment to Dworetsky's question: Fossil field
  advocates have always recognized that the cores of A stars will host
  dynamos! What this work does is to provide a detailed picture which,
  comfortingly, agrees reasonably well with prior estimates.  As the
  speaker said, the question is whether the field can manifest itself at
  the surface.  So there is no need for a change in paradigm.  Question: I
  assume that there is not a large dipole moment of the dynamo field?}

\discuss{Browning}{The maximum amplitude of the dipole component is
  typically about 5\% of the maximum amplitude of the total radial magnetic
  field.  So yes, the dipole is fairly weak.}

\end{discussion}

\end{document}